\documentclass[12pt,preprint]{aastex}
 
\shorttitle{Line Properties of Seyfert~1 Type AGN}
\shortauthors{D.W. Xu et al.}

\begin{document}

\title{An AGN Sample with High X-ray-to-optical Flux Ratio from RASS\\
II. Optical Emission Line Properties of Seyfert~1 Type AGN} 
\author{D. W. Xu\altaffilmark{1,2}, Stefanie Komossa\altaffilmark{2}, J. Y. Wei\altaffilmark{1}, Y. Qian\altaffilmark{3} and X.Z. Zheng\altaffilmark{1}}
\altaffiltext{1}{National Astronomical Observatories, Chinese Academy of
Sciences, Beijing 100012, China; dwxu@bao.ac.cn, wjy@bao.ac.cn, 
zxz@alpha.bao.ac.cn}
\altaffiltext{2}{Max-Plank-Institut f\"{u}r extraterrestrische Physik, Postfach
1312, D-85741 Garching, Germany; dwxu@mpe.mpg.de, skomossa@mpe.mpg.de}
\altaffiltext{3}{Department of Physics, Tsinghua University, Beijing 100080, 
China}

\begin{abstract}
This work studies the optical emission line properties of a sample of 155
low-redshift bright X-ray selected ROSAT Seyfert~1 type AGN for 
which adequate
signal-to-noise ratio spectroscopic observations are available.
We measured emission line properties by
performing multi-component fits to the emission line profiles, covering the
effect of blended iron emission. We also obtained continuum parameters, including
250eV X-ray luminosities derived from the ROSAT database. In addition, 
the measured properties are gathered for a correlation analysis, which confirms
the well-known relations between the strengths of Fe~II, [O~III] emission and
the X-ray slope.
We also detect striking correlations between H$\beta$
redshift (or blueshift), flux ratios of Fe~II to H$\beta$ broad component
and [O~III] to H$\beta$ narrow component. These trends are most likely driven
by the Eddington ratio. 
\end{abstract}


\section{Introduction}
Optical spectra of Active Galactic Nuclei (AGN) exhibit an extremely 
wide variety of properties.
The emission-line regions of AGN have proven to be among the most complex
of all astrophysical environments \citep{cb96}. 
AGN have also been known to emit a substantial fraction of their luminosity 
as X-rays. Multi-wavelength
observations of well-defined source samples are effective in
understanding the physical processes in AGN.  
In the last decade, several correlations between optical emission line 
properties and X-ray properties were noted \citep{bg92,law97,gru99,vau01}: 
for instance, the strong correlations between optical Fe~II, 
[O~III]$\lambda5007$ line strengths, velocity width of H$\beta$ and
the slope of soft X-ray continuum (${\rm \alpha_X}$). 
These correlations received great interest previously 
and were expected to give important insight 
into the AGN phenomenon. Still,
however, their interpretation remained unclear. It is therefore of great
importance to test whether these correlations are fundamental properties or
subtle selection effects, and search for new ones. Clearly, a large and
homogeneous X-ray selected sample would be advantageous in addressing 
these issues. We present such a sample here.

X-ray AGN samples are also of great interest for a number of other issues.
AGN of different types are the major contributors to 
the extragalactic soft and hard X-ray background (XRB). 
The definitions of statistically 
significant AGN samples using X-ray data are key to test AGN number-flux 
relations, luminosity functions and cosmological evolution. 
In the course of the ROSAT All Sky Survey 
(RASS) \citep{vog96,vog99}, many new AGN have been identified by 
different teams, confirming that X-ray surveys are very efficient in finding 
new AGN. \citet{sch00} performed an unbiased survey of all bright, 
${\rm CR>0.2~s^{-1}}$, high galactic latitude sources in the RASS without 
applying other selection criteria. 
\citet[Paper I]{wei99} identified a sample with high 
X-ray to optical flux ratio. \citet{app98} and \citet{zic97}
presented a catalog of sources in selected areas down to the RASS 
limit. \citet{tho98} drew a high galactic latitude sample from the RASS 
with soft PSPC spectra (${\rm HR1<0.0}$), the corresponding 
sample of point-like
X-ray sources with relatively hard X-ray spectra (${\rm HR1>0.5}$) 
was presented by 
\citet{fis98}. \citet{bade95,bade98} published a catalog of 
northern AGN in the RASS based on their Schmidt plate survey. 
These authors presented new AGN, but they did not use their samples
to address correlations issues. This will be done with the present sample.

In Paper~I we described and presented the optical identification of  
a sample of 165 X-ray sources with high X-ray to optical flux ratio 
(${\rm f_X/f_{opt}}$)
discovered in the RASS. About 73\% of the sources were identified as AGN,
including 115 emission line AGN (QSOs and Seyferts) as well as 2 BL Lac objects
and 4 BL Lac candidates. In this paper we investigate
the optical emission line properties of the emission line AGN 
that were identified as main 
optical counterparts of the sample and discuss their connection 
with X-ray properties. Not included in the following discussion are the BL 
Lac objects. 

The initial identification program excluded known X-ray sources 
that have counterparts in the SIMBAD, NED and AGN catalogs. 
Motivated by the need of larger and better samples, we have
cross-referenced the RASS bright source catalog with the NED and AGN catalogs. 
We find that 40 known emission line AGN meet our selection criteria.%
\footnote{The selection criteria of the sample are: an alternative high
X-ray-to-optical flux ratio criterion, i.e., 
${\rm \log CR \geq - 0.4\,R + 4.9}$,
where CR and R represent X-ray count rate and R magnitude respectively; 
declination
${\rm \delta \geq 3\degr}$; galactic latitude ${\rm |b| \geq 20\degr}$;
optical counterparts within a circle with radius ${\rm r=r_{1}+5\arcsec}$,
where $r_{1}$ is the RASS position error given by Voges~ et~ al.(1996);
optical counterparts with R magnitudes between 13.5 and 16.5.}
In order to 
keep consistent emission line properties measurements, new optical 
spectra have been acquired for the 40 additional sources. 
These data, combined with the previous optical 
identifications, provide a large, bright and (${\rm f_X/f_{opt}}$) limited 
sample of emission line AGN in 
the ROSAT X-ray band. The relatively complete sample is the largest X-ray
selected AGN sample with intense optical investigation so far.

As described in Paper~I, our classifications and identifications were based on 
low resolution spectra obtained with a 2.16~m telescope at Xinglong station,
National Astronomical Observatories of China (NAOC). These spectra were 
exposed to reach a S/N suitable for a reliable classification.  
The quality of the majority spectra is sufficient to measure the properties 
of the emission lines, such as the line widths and line strengths. For spectra
with inadequate S/N, new low resolution ($\sim$10\AA~FWHM) optical 
spectroscopy was performed using the 2.16~m telescope at NAOC. 

All objects in our emission line AGN subsample are found to be Seyfert 
1 type AGN. Throughout this paper 
Seyfert 1 galaxies and QSOs will be discussed jointly, since these two classes
are distinguished only by a certain luminosity level according to  our 
classification criterion. We do not divide the 
Seyfert 1 class into subclasses such as Seyfert 1.5-1.9 or NLS1 galaxies 
(as defined by \citet{ver00}, \cite{ost85}), 
since the definitions of Seyfert classifications depend on 
the resolution of the spectra used and the noise in the spectra \citep{gru99,
good89a,good89b}.

The paper is organized as follows. Detailed optical measurements are 
given in Sect.~2. The statistical properties are investigated in Sect.~3. 
Several diagnostically important correlations that exist between emission 
line and continuum properties are studied. 
Sect.~4 discusses the implications of the results, followed by a brief
summary of the main conclusions of the paper.

Throughout this paper, a Hubble constant of ${\rm H_{0}=50~km~s^{-1}~Mpc^{-1}}$
and a deceleration parameter of ${\rm q_{0}=0}$ are assumed and all 
measured parameters are quoted for the rest frame of the source.

\section{Observations and Optical Measurements}

\subsection{New Optical Observations}
The known sources as well as sources without high quality observations in the
identification campaign were investigated during several observing 
runs performed from November 1998 through December 2000. 
The low resolution spectra were taken
with the NAOC 2.16~m telescope and the OMR spectrograph, using a 
Tektronix $1024 \times 1024$ CCD as detector.
The grating of 300~g~mm$^{-1}$ was employed in order to get large 
wavelength coverage. All observations were made through a 2.3$''$ slit
that produced a resolution of 10 to 11\AA~as measured from the comparison
spectra. This set-up provided similar wavelength coverage
(3800\AA~--~8200\AA) and resolution to
the identification data. Exposure time was generally between 1200s and 
3600s depending on the brightness of the object.  For a few objects, 
multiple exposures were performed in order to get a higher S/N spectrum.
The exposures were combined prior to extraction where possible in order to help
remove cosmic ray contamination. In cases when the target source moved across 
the CCD chip between exposures, the data were extracted separately from each
frame and then combined.

The raw data were reduced following standard procedures using IRAF. 
The CCD reductions included
bias subtraction, flatfield correction and cosmic-ray removal.
Wavelength calibration was carried out using helium-neon-argon lamps
taken at the beginning and end of each exposure. The resulting
wavelength accuracy is better than 1~\AA. The flux calibration was 
derived with 2 to 3 observations
of KPNO standard stars \citep{mas88} per night.

\subsection{Fe~II Subtraction} 
In many of the spectra there is a clear contribution from blends of 
Fe~II line emission on both the blue and red sides of the H$\beta$-[O~III] 
complex. These blends contaminate strong lines such as 
[O~III]$\lambda\lambda$4959,5007, and may alter the flux and width of 
H$\beta$+[O~III]$\lambda\lambda$4959,5007. In order to reliably 
measure line parameters and
to determine the strength of the Fe II emission, we have carefully
removed the Fe~II multiplets following the method described by
\citet{bg92}, which relies on an Fe~II template.
The template used in the present work 
is the same as that of \citet{bg92}, namely the Fe~II lines of I~Zw~1, a bright NLS1
widely known for the strong and narrow permitted Fe~II emission \citep{phi78,
oke79}. The observed spectra were wavelength-shifted to the 
rest frame according to the redshift.%
\footnote{We take the redshift from [O~III]$\lambda$5007 to be the systemic 
redshift as most commonly used. The redshift was derived from fitting a 
Gaussian to
the upper half of the observed [O~III]$\lambda$5007 line in each spectrum.}
The whole template was broadened to
the FWHM of the broad H$\beta$ line by convolving with a Gaussian and
scaled to match the line intensities. The best match was then searched for
in the two-dimensional parameter space of line width and line strength.
From each object the best-fit Fe~II spectrum was then subtracted.  A successful 
Fe~II subtraction showed a flat continuum between the H$\gamma$ and
H$\beta$ and between 5100--5400\AA~(which covers the Fe~II multiplets 48, 49).
The Fe~II flux was measured between the rest wavelength 4434\AA~and 4684\AA~ as 
in \citet{bg92}.

\subsection{Other Line Measurements}
The characteristic property of Seyfert~1 type AGN  
is the presence of broad emission lines. The line 
widths and their distributions provide important information on masses of the 
central black holes and mass distributions. 
The broad Balmer lines in AGN exhibit a wide variety of profile shapes and a 
large range in width \citep{ost82,de85,cre86,str91,mil92,ver01} and they 
are often 
strongly asymmetric \citep{cor95}. In many cases H$\beta$ is a mixture of broad 
and narrow components. Differences in the relative strengths of these 
components account for much of the diversity of broad line profiles 
\citep{fra92,wil93,bro94,cor95,cor97,ver01}.

The Fe~II subtracted spectra were used to measure the non-Fe~II line properties.
In order to isolate the broad H$\beta$ component in all spectra, we have 
assumed that the emission line profiles 
can be represented by a single or a combination of 
Gaussian profiles. \citet{ver01} claimed that the broad emission lines in 
NLS1 galaxies were better fitted with Lorentzian profiles than with 
Gaussian profiles.
However, as noted in \citet{eva88}, the choice of Gaussian or Lorentzian
profiles as representatives of the observed emission lines may bear no physical 
meaning. We performed a multiple Gaussian fit 
in the present work for its simplicity and a direct comparison
with the measurements of previous studies (e.g. \citet{bg92, gru99, vau01}).
We adopted the method described by \citet{rod00b} and used the IRAF 
package SPECFIT%
\footnote{SPECFIT is developed and kindly provided by Dr. G. Kriss.}
to measure blended lines.
As a first step we tried to
fit the emission lines with a single Gaussian component. 
It worked well for most forbidden lines, e.g., [O~III]$\lambda$5007.
However, this simplest representation 
could not fit adequately the wings of H$\beta$, 
although its core was nicely fitted in most of targets. In such cases, we then 
included one narrow component and one broad component 
to describe adequately the observed profile. (Note that when 
H$\beta$ measurements were taken from the literature, many of them 
refer to the whole line.) For each assumed component, 
a first guess of the
central wavelength, the total flux, and width of that component were specified.
There were two parameters for each continuum component, i.e., the flux and the
slope. Initially, we constructed a narrow component with a width determined from
fitting the [O~III] lines. In some spectra such a guess failed to give
a good fit and needed an additional component. We then left the width of
the narrow component as a free parameter. It led to a nice fitting result
by checking the residual, i.e., the narrow component could be well isolated 
from the broad component. For H$\beta$ lines with asymmetric profile, the
two Gaussian components are not stationary in wavelength. 
The velocity shift between the broad and narrow components is defined as 
the shift of the broad component relative to the
narrow component in units of km~s$^{-1}$ (see also Zheng et al 2002). 
As an illustration, we show the decomposition in Figure~1. The observed
profile is represented by a solid line, the total fit by a dashed line,
the individual Gaussian components by a dot-dashed line and the
differences between the data and the fit (the residual) by the lower
dotted line.

The broad component is referred as H$\beta_{b}$, and the narrow component
as H$\beta_n$. The FWHM H$\beta_{b}$ is what we consider as 
representative of the Broad Line Region (BLR). The instrumental resolution 
was determined from the FWHMs of night sky spectra taken with each of the 
exposures and tested by comparison-lamp spectra. 
All quoted line widths were corrected for the instrumental resolution. 
The uncertainty of
H$\beta$ velocity shift measurement is within 150~km~s$^{-1}$.
To measure line fluxes and equivalent widths (EW), power-law continua 
were fitted and subtracted from underneath the lines and the remaining 
line flux was integrated. For most objects, the uncertainty of flux
measurement of emission lines is about 10\% to 15\%.

\subsection{Continuum Measurements}
The continuum level at the position of H$\beta$ $\lambda$4861 was 
measured from the 
Fe~II subtracted spectra. All equivalent widths measurements in this paper refer
to the continuum at that point to allow direct comparison with the measurements
for other samples \citep{bg92,gru99,vau01}.
The optical index ${\rm \alpha_{opt}}$ was calculated using the continuum 
flux density at 4000\AA~ and 7000\AA~ in the rest frame. 

For the X-ray analysis, a single power law description was employed. This
application has proven to be a fairly good approximation for AGN in
previous studies 
(e.g. Fiore et al. 1994, Brinkmann \& Siebert 1994, Schartel
et al. 1996, Grupe et al. 1998). Because of the limited 
signal-to-noise ratio of the RASS spectra
we assumed that the absorption comes from our Galaxy only. The assumption
usually improves the reliability of the estimation of spectral index because
of the reduced number of free parameters. The X-ray spectral slope 
${\rm \alpha_{X} (F_\nu \propto \nu^{-\alpha})}$ was 
estimated using the ROSAT hardness ratios HR1 and HR2, which are defined
as \citep{vog99}:
$${\rm HR1=\frac{B-A}{B+A}}$$
$${\rm HR2=\frac{D-C}{D+C}}$$
Here, A, B, C and D denote the count rate in the energy range 0.1--0.4 keV,
0.5--2.0 keV, 0.5--0.9 keV and 0.9--2.0 keV, respectively.
 
\section{Analysis}
\subsection{General Sample Properties}
The full optical dataset was used to define the 
statistical properties of the sample
and explore the correlations between various observed parameters.
According to Paper~I, 115 sources (about 73\% of the total) were identified
as new Seyfert~1s and QSOs, together with 40 known emission line AGN that
meet our sample selection criteria,
the AGN subsample comprises 49 ``ultrasoft'' 
($\alpha_{X}>1.7$) sources and 106 ``normal'' objects
with mean $<\alpha_{X}>=1.43$. In the catalog in Paper~I we listed the 
absolute magnitude M$_{B}$ and redshift $z$. The distributions of the M$_{B}$
values and redshift are plotted in Figure~2. 
As shown by these figures, the sample is dominated by moderate luminosity 
objects (${\rm M_B \approx -23 \pm2}$) with an average redshift ${\rm <z>=0.2}$.

Table~1 lists the derived parameters. The columns list the following
information: (2) redshift; (3) FWHM of broad H$\beta$; 
(4)  FWHM of narrow H$\beta$;
(5) FWHM of [O~III]$\lambda$5007; (6), (7) and (8) the local equivalent
widths of H$\beta$ broad component, Fe~II and [O~III]$\lambda$5007;
(9) and (10) ratio of fluxes of [O~III]$\lambda$5007 to broad and narrow
H$\beta$;
(11) ratio of fluxes of  Fe~II  to broad H$\beta$;
(12) velocity shift
between H$\beta$ broad and narrow components, where 
a positive velocity corresponds
to a redshifted broad component relative to narrow component;
(13) and (14) H$\beta$ broad and narrow components velocity shift
with respect to the systemic frame, where a positive velocity
indicates a redward shift with respect to systemic;
(15) and (16) spectral indices defined above;
(17) the monochromatic 250eV luminosity.

The distribution of the broad H$\beta$ line width of the objects in our 
emission line AGN subsample shows a 
significantly smaller fraction (3\%) of objects with FWHM $<$ 2000
~km~s$^{-1}$ compared to another X-ray selected sample (18\%)
\citep{app00}. However,  
the FWHM measurements in \citet{app00} refer to the full line
profile and the widths of the broad components may have been underestimated.
Therefore, the real difference should be smaller. Five out of 155 objects in
our sample are
NLS1 galaxies without detectable broad Balmer line components
of FWHM $>$ 2000~km~s$^{-1}$ but with strong Fe~II emission,
which agrees within the error limits with that of 
\citet{app00} (about 2\%). 
The distribution of the
FWHM of the broad component of H$\beta$ in the sample is displayed
in Figure~3.  The 
mean value of 4350 km~s$^{-1}$ is similar to the hard sample of 
\citet{gru99}.

\subsection{Distribution of Velocity Shifts}
The velocity shift distribution of H$\beta_b$ with respect 
to [O~III]$\lambda5007$ is shown in Figure~4a. Positive velocity corresponds 
to a redward  H$\beta$
with respect to systemic, while a negative velocity represents a blueshift.
The peak of this distribution is near zero, but has a mean redshift of 
78~km~s$^{-1}$ with standard deviation of 305~km~s$^{-1}$. This shift 
is in good agreement with the previous studies \citep{gaskell82,tytler92,
bg92,laor95,cb96}, i.e., Balmer emission lines gives the same redshift to 
within 100~km~s$^{-1}$ of narrow forbidden emission ({\it e.g.} 
[O~III]$\lambda$5007) in low redshift ($z<1$) QSOs and Seyfert galaxies. In
contrast, a systematic mean redward shift of 520~km~s$^{-1}$ was found 
for a high redshift ($2.0 \lesssim z \lesssim 2.5$) QSOs sample 
\citep{mci99}. \citet{mci99} proposed this observed trend of increased Balmer
redshift with increased systemic redshift represents a luminosity dependency.

For comparison, we also present the H$\beta_n$ -- [O~III]$\lambda5007$ shift
in Figure~4b. The mean of the distribution is 8~km~s$^{-1}$ with a standard
deviation of 176~km~s$^{-1}$. We can see from the figures that the dispersion
of the H$\beta_n$ -- [O~III]$\lambda5007$ shift is significantly narrower than
the H$\beta_b$ -- [O~III]$\lambda5007$ shift, which is expected by the different
emitting regions between H$\beta$ broad component and narrow component,
as well as [O~III]$\lambda5007$.
 
In a similar manner to the H$\beta_b$ -- [O~III]$\lambda5007$ shift,
Figure~4c illustrates the velocity differences between the broad and 
narrow components of H$\beta$. The redshift (or blueshift) is defined
as the shift of the broad component relative to the narrow component
in units of km~s$^{-1}$. Positive velocities refer to redshifts 
with respect to the narrow component velocities, whereas negative
velocities indicate blueshifts. We emphasis that both the H$\beta$ redshift 
(or blueshift) in the present work and the asymmetry parameter in \citet{bg92}
are measures for line profile, but characterized differently. The latter is
a measure of the shift between the centroids at $1/4$ and $3/4$ maximum. 
The mean redshift of the distribution
is found to be 70~km~s$^{-1}$ with a standard deviation of 357~km~s$^{-1}$.
Correlation analyses involving the line profiles of H$\beta$ are 
investigated in section 3.3.  

\subsection{Correlations Analysis}
In this subsection, we explore whether the various emission-line and 
continuum properties correlate with one another. For this purpose,
we calculated the Spearman rank-order correlation matrix \citep{pre92},
along with its significance matrix for measured properties.
The complete correlation coefficient matrix is shown in Table~2. 
The number of parameter pairs included in the trials 
ranges from 86 to 149. Spectra with S/N $<10$ were excluded from the 
analyses including optical properties. The probability of the null 
correlation, Ps, for a sample with corresponding correlation coefficient 
Rs is also given in Table~2 for entries with ${\rm Ps<0.01}$.
A set of 12 different
properties results in $12 \times (11/2)=66$ correlation coefficients; therefore,
at this level of significance, we would expect $\lesssim 1$ spurious events.

Among 66 trials, 31 correlations (both positive and negative) are found with
two-sided probabilities ${\rm Ps<0.01}$. Four are due to the 
dependent parameters, e.g., relative strength of 
Fe~II to H$\beta$ and Fe~II~EW; 12 are degenerate correlations, i.e.,
different measures of the same property (e.g., [O~III]~EW, [O~III]/H$\beta_b$ 
and [O~III]/H$\beta_n$) all correlate with another property (e.g., Fe~II~EW);
and therefore 15 are independent correlations at the $\gtrsim 99\%$ confidence
level. 

The well-known anti-correlation between the X-ray spectral slope 
${\rm \alpha_{X}}$ and the FWHM of H$\beta$ \citep{bol96,lao97,gru99} is not
prominent in our sample (${\rm Rs=-0.20}$, ${\rm Ps=0.05}$) 
as is the case in the \citet{vau01} sample 
(${\rm Rs=-0.31}$, ${\rm Ps=0.03}$). \citet{vau01} suggested 
that the relation between H$\beta$ width and ${\rm \alpha_{X}}$ is not 
a linear correlation, but probably resulted from a ``zone of avoidance'' 
\citep{kom01}, i.e., 
broad line objects always tend to show flat X-ray spectra, whereas there 
is a very large scatter in the X-ray spectral steepness of NLS1 galaxies, 
with several as flat as normal Seyfert 1s (Xu et al. 1999). 
The correlation is more pronounced among the AGN with higher luminosity 
as in the samples of 
\citet{gru99} and \citet{vau01}. For objects with ${\rm \nu L_{250eV} >
10^{44.3}~erg~s^{-1}}$ the significance becomes ${\rm Rs=-0.45 (Ps=0.007)}$.
The correlation between ${\rm \alpha_{X}}$ and  H$\beta_b$ is illustrated in
Figure~5.

The anti-correlation between the width of H$\beta$ and Fe~II/H$\beta$ 
\citep{wil82,bg92,wan96,rod00a} is strong (${\rm Rs=-0.59}$, 
${\rm Ps<10^{-4}}$) in the 
present data whereas it is only weak (${\rm Rs=-0.31}$, ${\rm Ps=0.002}$) if 
FWHM H$\beta$ 
is compared directly with EW Fe~II. Instead, there is a strong correlation
(${\rm Rs=0.56}$, ${\rm Ps<10^{-4}}$) between FWHM H$\beta$ and EW H$\beta$. 
The lack of correlation between 
the line width and Fe~II strength (when measured independently with H$\beta$)
in \citet{gru99}, \citet{vau01} and the present work has confirmed
that the anti-correlation arises because H$\beta$ gets weaker
as the lines get narrower \citep{ost77,gas85,good89b,gas00}.  

In addition to our confirmation of well documented correlations between
different AGN properties, we report on three newly discovered correlations
involving the H$\beta$ redshift (or blueshift), flux ratios of Fe~II to 
H$\beta$ broad component (Fe~II/H$\beta_b$) and [O~III] to H$\beta$
narrow component ([O~III]/H$\beta_n$). The correlations covering 
[O~III]/H$\beta_n$ have never been investigated by other authors so far. 
We find the relative strength of Fe~II to H$\beta_b$
(Fe~II/H$\beta_b$) significantly correlated to both the
H$\beta$ redshift%
\footnote{A correlation between the H$\beta$ blueshift and flux ratio of Fe~II
to H$\beta$ whole line was detected in \citet{zhe02}. We emphasize that
our ratio refers to the broad component in H$\beta$ only. Note the sign 
difference between this paper and \citet{zhe02}.
The positive correlation quoted by \citet{zhe02} becomes negative in 
the notation used herein.} 
(${\rm Rs=-0.64}$, ${\rm Ps<10^{-4}}$)
and the relative strength of [O~III]
to H$\beta_n$ ([O~III]/H$\beta_n$, ${\rm Rs=-0.60}$, ${\rm Ps<10^{-4}}$). 
A correlation test was also applied between H$\beta$ redshift and 
[O~III]/H$\beta_n$. The correlation is prominent in our sample
(${\rm Rs=-0.38}$, ${\rm Ps<3 \times 10^{-4}}$). 
Figure~6 shows the correlations.
There is a trend that strong Fe~II/H$\beta_b$--weak [O~III]/H$\beta_n$
objects tend to have blueshifts in H$\beta$ while strong [O~III]/H$\beta_n$--
weak Fe~II/H$\beta_b$ objects tend to have redshifts in H$\beta$. 
These strong correlations between
Fe~II/H$\beta_b$, H$\beta$ redshift and [O~III]/H$\beta_n$ must reflect 
some physical connection between broad and narrow line emitting regions. 

\section{Discussion}
One key to understand the central engines of AGN lies in examining their
local environment. The standard paradigm proposes that the surrounding gas
is photoionized in physically distinct regions by radiation emerging
from the central power source\citep{ost89}. However, the segregation of the
ionized gas into broad-line region (BLR) and narrow-line region (NLR), as well
as the details of the gas kinematics are still far from clear.
Optical spectroscopy of the ionized gas around galactic
nuclei provides strong constraints on the excitation mechanisms.
The following discussion addresses three key points: (1) how do the 
present results compare to previous work; (2) which new correlations
do we find and how can we understand them; and (3) which future 
observations can we perform to distinguish between different 
suggested scenarios.
 
A number of consistent correlations among observational parameters
have been searched in order to understand the basic
properties underlying the observed spectra. In particular,
\citet{bg92} made a landmark study  of the optical emission-line
properties and continuum properties (radio through X-ray) of
87 low redshift PG quasars. A principal component analysis revealed
``eigenvector~1'' (EV1) links stronger Fe~II emission, weaker
[O~III] emission from the NLR,  and narrower H$\beta$ (BLR) with 
stronger line blue asymmetry. More recently, \citet{lao97} found these optical
properties also go along with steeper soft X-ray spectra and claimed
the soft X-ray slope as a related part of EV1.
We confirm these trends and present in addition three 
newly discovered correlations involving the H$\beta$ redshift, 
Fe~II/H$\beta_b$ and [O~III]/H$\beta_n$.

Potentially, EV1 represents a fundamental physical driver that
control the energy producing and radiation emitting processes.
The two leading interpretations by far are that EV1 is driven
by (1) the Eddington ratio $L/L_{Edd}$ \citep{bg92} and (2) Orientation effect.
 
\citet{bg92} argued strongly that viewing angle is
unlikely to drive EV1, by assuming [O~III] emission is isotropic,
originating from radii large enough to be free from orientation dependent
obscuration. The isotropy of [O~III] emission has been questioned 
by the studies of radio-loud AGN \citep{hes93,bak97}. However, a recent study
of radio-quiet quasars \citep{kur00} showed a significant
correlation between EV1 and orientation independent [O~II] emission, which
implied that EV1 is not driven by orientation. Moreover, the correlation
between [O~II] and [O~III] emission indicated [O~III] emission is not
dependent on orientation.

The most promising interpretation at present is that EV1 is mostly
governed by $L/L_{Edd}$ \citep{bg92}. This suggestion was based on
the notion that the vertical thickness of the accretion disk, driven
by the Eddington ratio, controls the line strengths and continuum parameters.
 
The striking correlations between the H$\beta$ redshift,
Fe~II/H$\beta_b$ and [O~III]/H$\beta_n$ found in the present work
give us linkages of the gas kinematics and ionization between BLR and NLR.
The question raised is what is the physical process behind the
correlations? Can both the new and previous correlations be explained within
one single scenario? Indeed, the $L/L_{Edd}$ 
interpretation provides plausible explanations for the observed 
trends.

One of the two general scenarios in AGN to explain asymmetric profile is
in terms of an outflowing component in the BLR with a change in
obscuration by a central disk of clouds on the receding side. Accreting
systems in high $L/L_{Edd}$ AGN will drive relatively thick outflows
due to their larger photon luminosities per unit gravitational mass.
As the velocity of the outflow increases, the H$\beta$ develops 
an increasing excess on the blue wing.
The corresponding blue H$\beta_b$ velocity shift in comparison with [O~III]
can be also well predicted because both the line asymmetry and the shift
of the line centroids are effects of the same process.
The link between BLR and NLR in terms of density is less straight forward
to predict. E.g., 
due to the strong outflow, the NLR
might be replaced by denser clouds. Given a typical  
radial distribution of NLR clouds, [O~III]/H$\beta_n$ peaks
around densities of $\log n$ = 2--4 (e.g., Komossa \& Schulz 1997).
High-density NLR clouds would therefore lead to a suppression of 
[O~III] strength. 
The same is true for the other limit, a low-density NLR. Some
authors favored a low-density BLR of NLS1 galaxies
(see, e.g., the discussion in Rodriguez-Pascual et al. 1997 
and Komossa \& Mathur 2001), which may indicate
a low-density NLR as well, if both regions are coupled.
The best way to measure directly the density of the
NLR of NLS1s in the future is to employ the density-sensitive line ratios
[O~II]$\lambda$3729/[O~II]$\lambda$3726 or 
[S~II]$\lambda$6716/[S~II]$\lambda$6731. 
These ratios are presently generally not reported in the literature. 
Their measurements requires high S/N spectroscopy and/or resolution
to detect and resolve these lines.%
\footnote{We checked the NLR density of the well-measured
NLS1 galaxy NGC 4051, using the density-sensitive [S~II]-ratio
measured by \citet{vei91}. This yields $\log n \simeq$ 2.8
and does not deviate significantly from the values found for, e.g.,
Seyfert 1 and 2 galaxies. However, NGC 4051 is not an
extreme NLS1 in terms of X-ray steepness and Fe~II emission.}
More indirectly, other line-ratios will also change in dependence of
density and thus allow a density determination. 
For instance, a higher-density NLR would strongly boost 
the [O~I]$\lambda$6300 line \citep{kom97}.  

Although it is more difficult to illustrate
the direct connection between strong Fe~II emission and
large $L/L_{Edd}$, some suggestions have been proposed.
\citet{pou95} suggested the X-ray spectrum  becomes steeper
in high $L/L_{Edd}$ AGN. If the EUV--X-ray spectrum plays an important role
in the formation and confinement of the 
BLR clouds (e.g., Krolik, McKee \& Tarter 1981), a steep X-ray
spectrum coupled to a flatter EUV spectrum
will lead to a thicker BLR (e.g., Komossa \& Meerschweinchen 2000), 
and thus stronger Fe~II emission if the
the low-excitation part of the BLR is mechanically heated \citep{jol87}.
Increased metal abundances in more extreme NLS1 galaxies 
\citep{mat00} would enhance these trends, i.e. lead to thicker
BLRs \citep{kom01}. Supersolar iron abundance
may additionally boost the strength of Fe~II. 
In another model by \citet{kwa95}, Fe~II line emission is produced
in an accretion disk. The AGN with higher $L/L_{Edd}$ might simply have
more mass in the accretion disk to produce stronger Fe~II emission by 
postulating collisional excitation \citep{net83,wil85} as the origin 
of the Fe~II emission in general.

The strong correlations of properties linking dynamics (line widths), 
kinematics (line shifts), ionization in BLR and NLR (Fe~II and [O~III] 
strengths), and radiation process (X-ray spectral shape) are likely
driven by an intrinsic properties, i.e., the $L/L_{Edd}$. 
In order to further test the favored scenario (and get more constraints
on, e.g., the density of the NLR), larger and even more 
complete AGN sample with higher resolution optical data is needed.

\section{Summary and Conclusions}
This paper presents the statistical study of optical emission line
properties of a bright AGN sample selected from RASS. The sample comprises
155 Seyfert~1 type AGN. Our conclusions based on the correlation analysis
can be succinctly summarized as follows:

\begin{itemize}
\item The previously reported correlations between the strengths of 
Fe~II, [O~III] emission
and the X-ray slope are confirmed. Strong optical Fe~II blends go along with
weak forbidden emission lines (e.g., [O~III]) and steep X-ray slopes.
\item The anti-correlation between the X-ray spectral slope
$\alpha_{X}$ and the FWHM of H$\beta$ becomes prominent as far as the
high luminosity AGN are concerned.
\item Striking correlations between the H$\beta$ redshift, Fe~II/H$\beta_b$
and [O~III]/H$\beta_n$ are found. There is a general trend that
strong Fe~II/H$\beta_b$--weak [O~III]/H$\beta_n$
objects tend to have blueshifts in H$\beta$ while strong [O~III]/H$\beta_n$--
weak Fe~II/H$\beta_b$ objects tend to have redshifts in H$\beta$. 
\item We add new arguments that the observed trends are likely driven by
$L/L_{Edd}$ and describe some future observations that could further test
this suggestion.   
\end{itemize}

\acknowledgments

We thank Drs. Todd A. Boroson and Richard F. Green for providing the 
Fe~II template, Drs. Dirk Grupe, Simon Vaughan, Mira V\'eron, Philippe V\'eron, 
Tinggui Wang, Dieter Engels for helpful discussions on the optical 
measurements, and the anonymous referee for many useful comments and 
suggestions. Special thanks go also to the staff at Xinglong station 
for their instrumental and observing help. D.X. acknowledges a MPG-CAS exchange
program. This research has made use of the 
NASA/IPAC Extragalactic Database (NED), which is operated by the Jet Propulsion 
Laboratory, Caltech, under contract with the National Aeronautics and Space 
Administration. The ROSAT project is supported by the Bundesministerium 
f\"{u}r Bildung und Forschung (BMBF) and the Max-Planck-Gesellschaft. This work
was supported by NSFC under grant 19973014.

\clearpage
\begin{figure}
\plotone{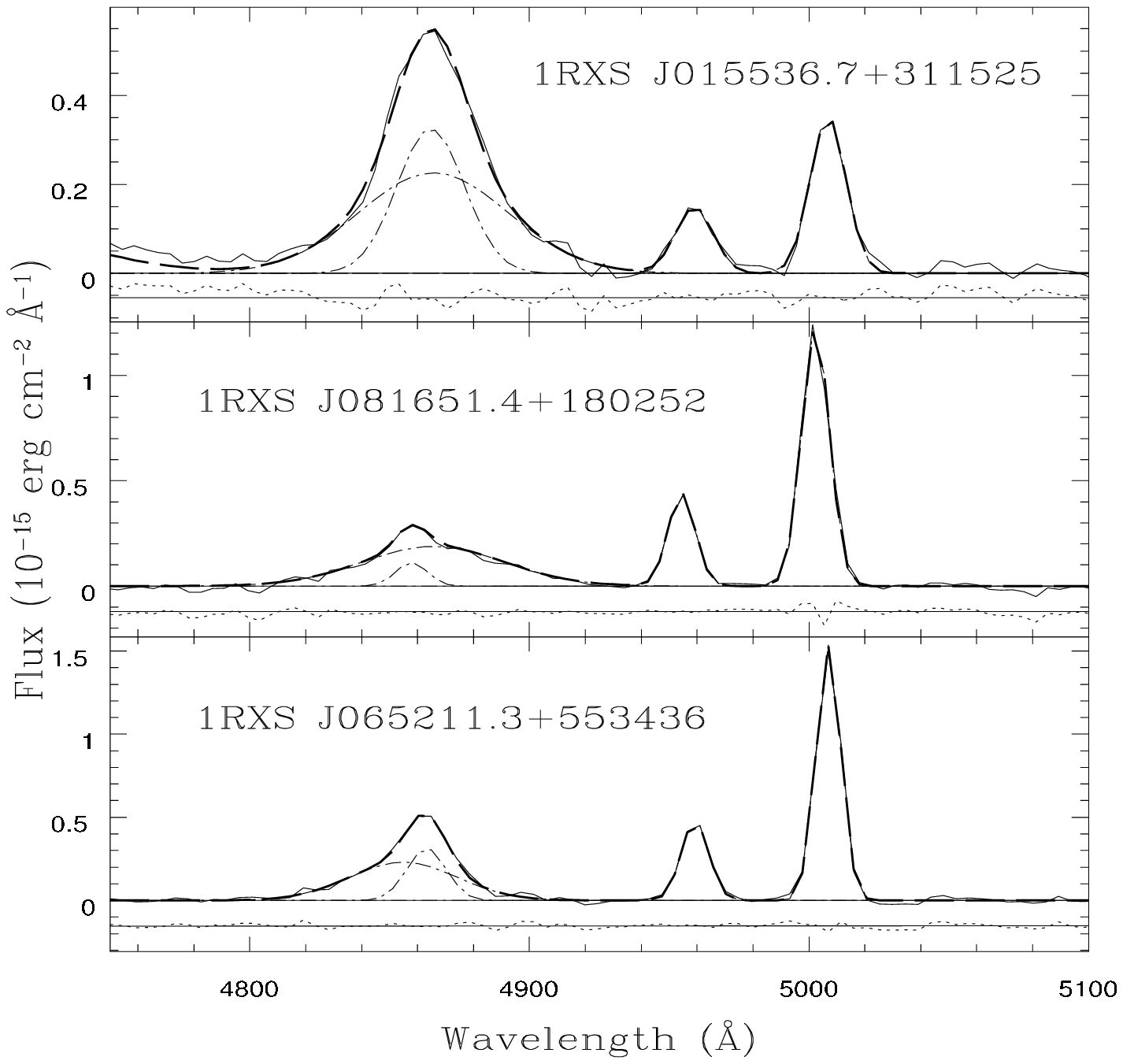}
\caption{Illustration of velocity shift between the broad and narrow
Gaussian components of H$\beta$. Top panel: For 1RXS~J015536.7+311525,
H$\beta$ can be well fitted by a sum of two Gaussian components without
velocity difference. Middle panel: For 1RXS~J081651.4+180252, H$\beta$
shows a redshifted broad component relative to the narrow one. Bottom
panel: For 1RXS~J065211.3+553436, H$\beta$ can be represented by
a blueshifted broad component and a narrow component.
 \label{fig1}}
\end{figure}

\clearpage
\begin{figure}
\plottwo{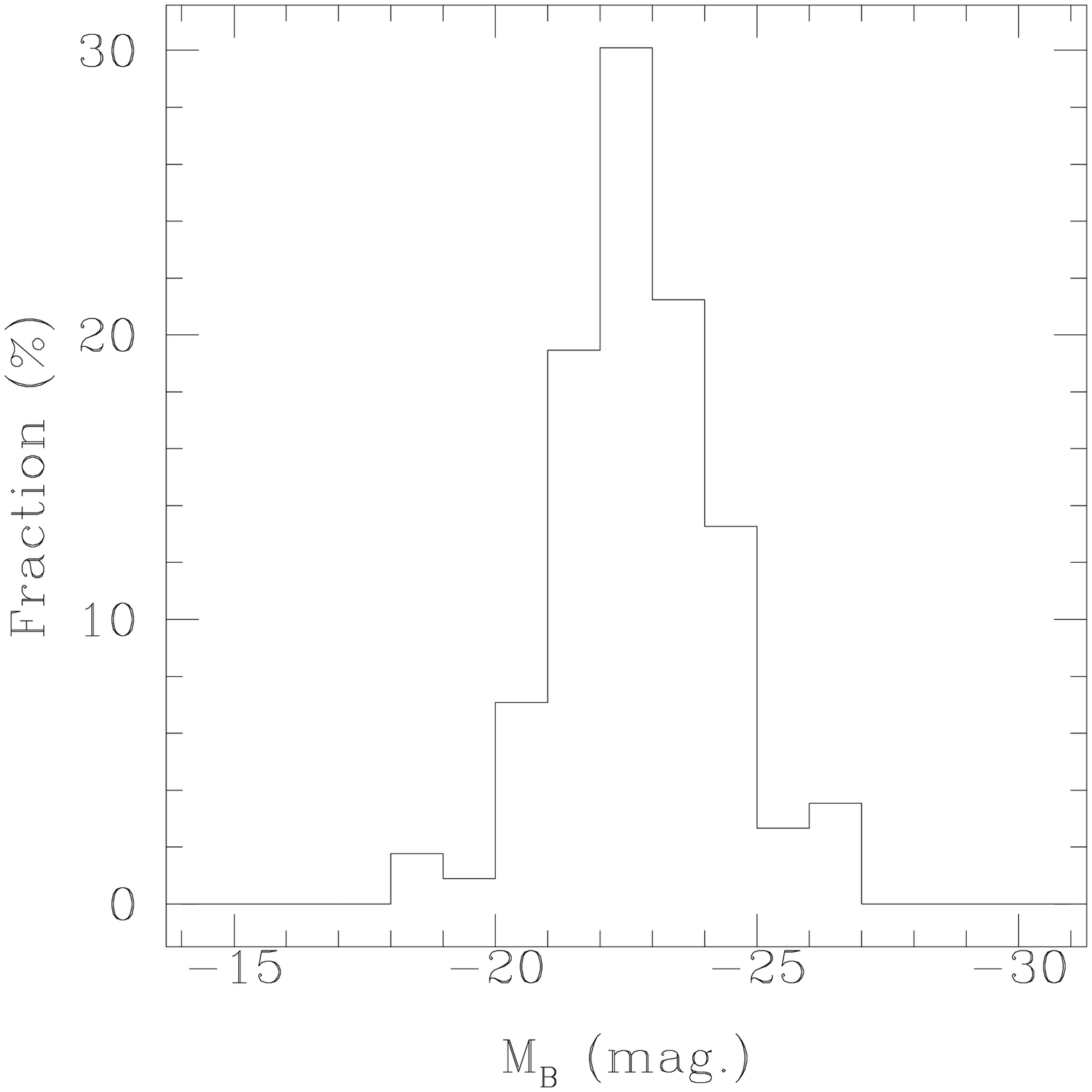}{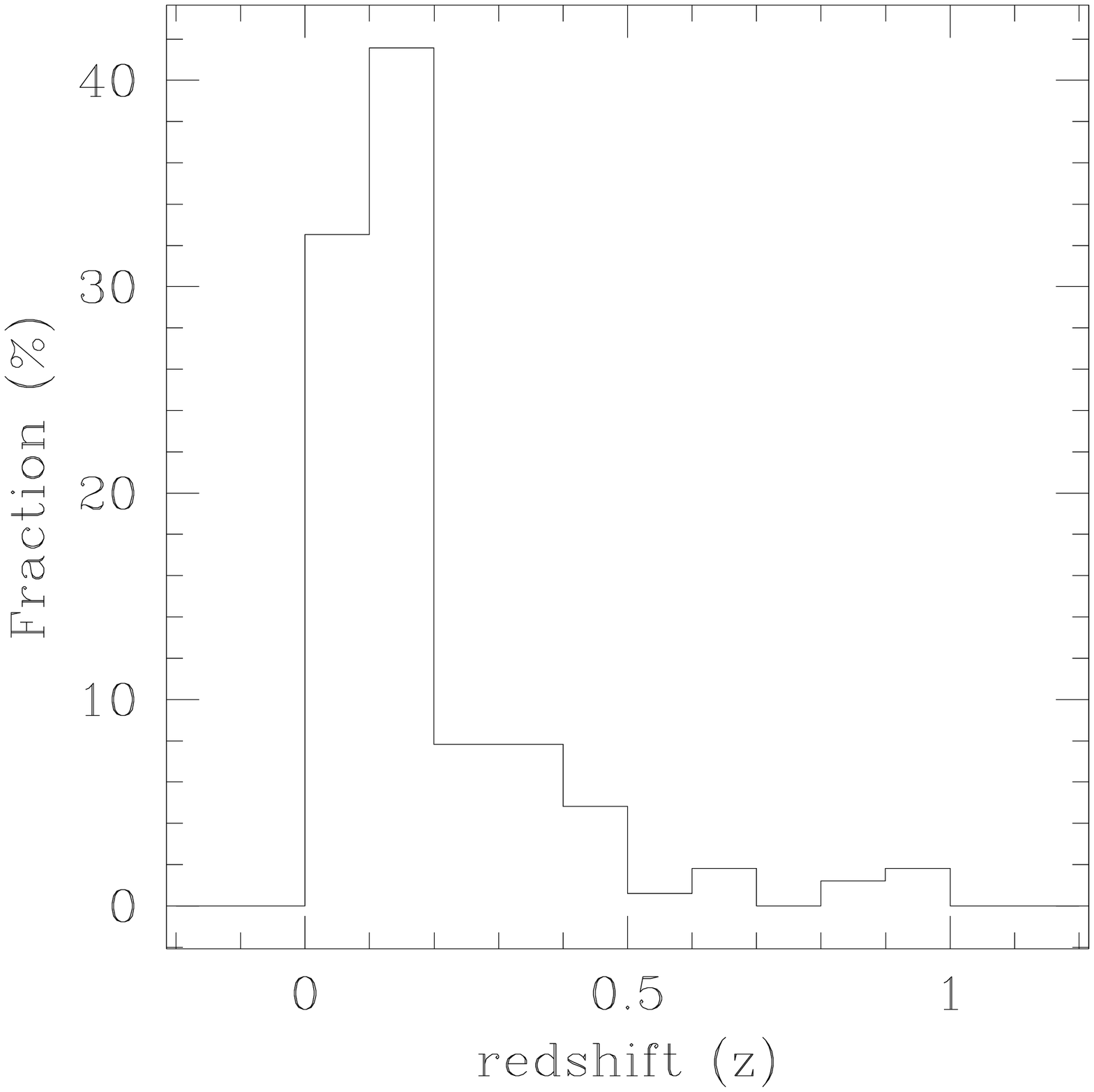}
\caption{Absolute magnitude M$_B$ and redshift distributions of the 
Seyfert~1 type AGN
\label{fig2}}
\end{figure}

\clearpage
\begin{figure}
\plotone{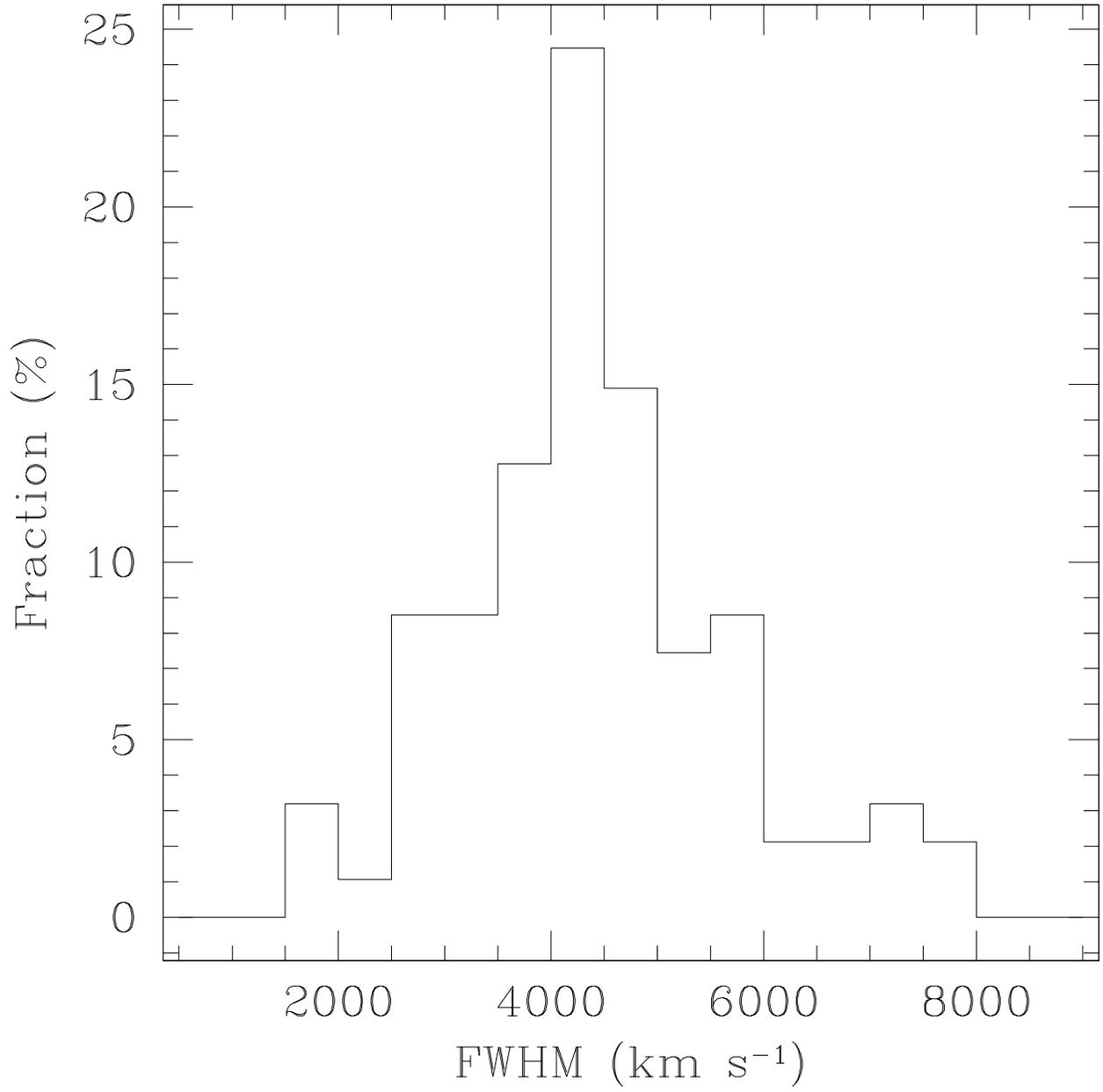}
\caption{Distribution of the FWHM of H$\beta$ for the Seyfert~1 type AGN
\label{fig3}}
\end{figure}

\clearpage
\begin{figure}
\plotone{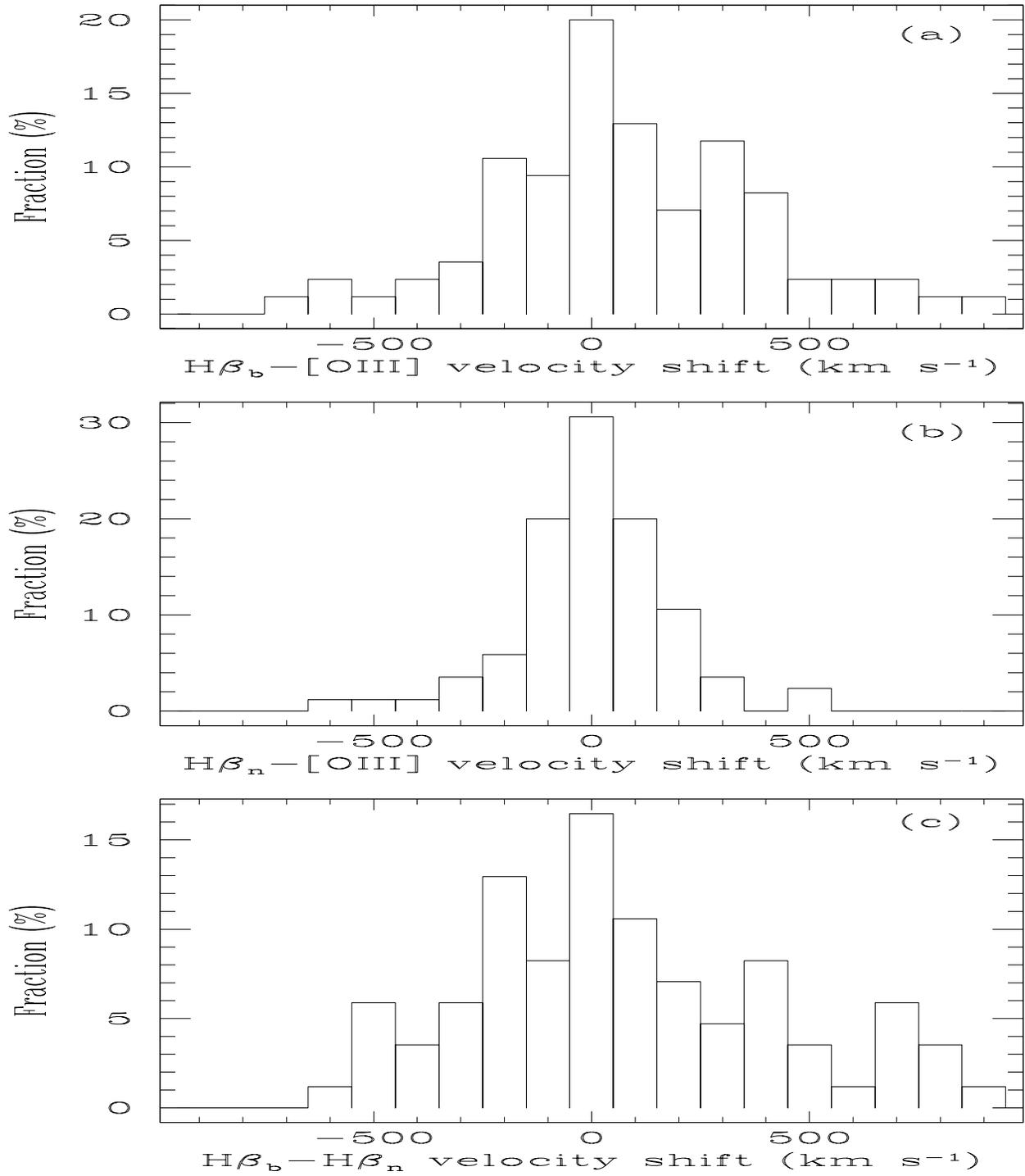}
\caption{Distribution of velocity shifts for H$\beta_b$,  H$\beta_n$ and
[O~III]$\lambda$5007
\label{fig4}}
\end{figure}

\clearpage
\begin{figure}
\plotone{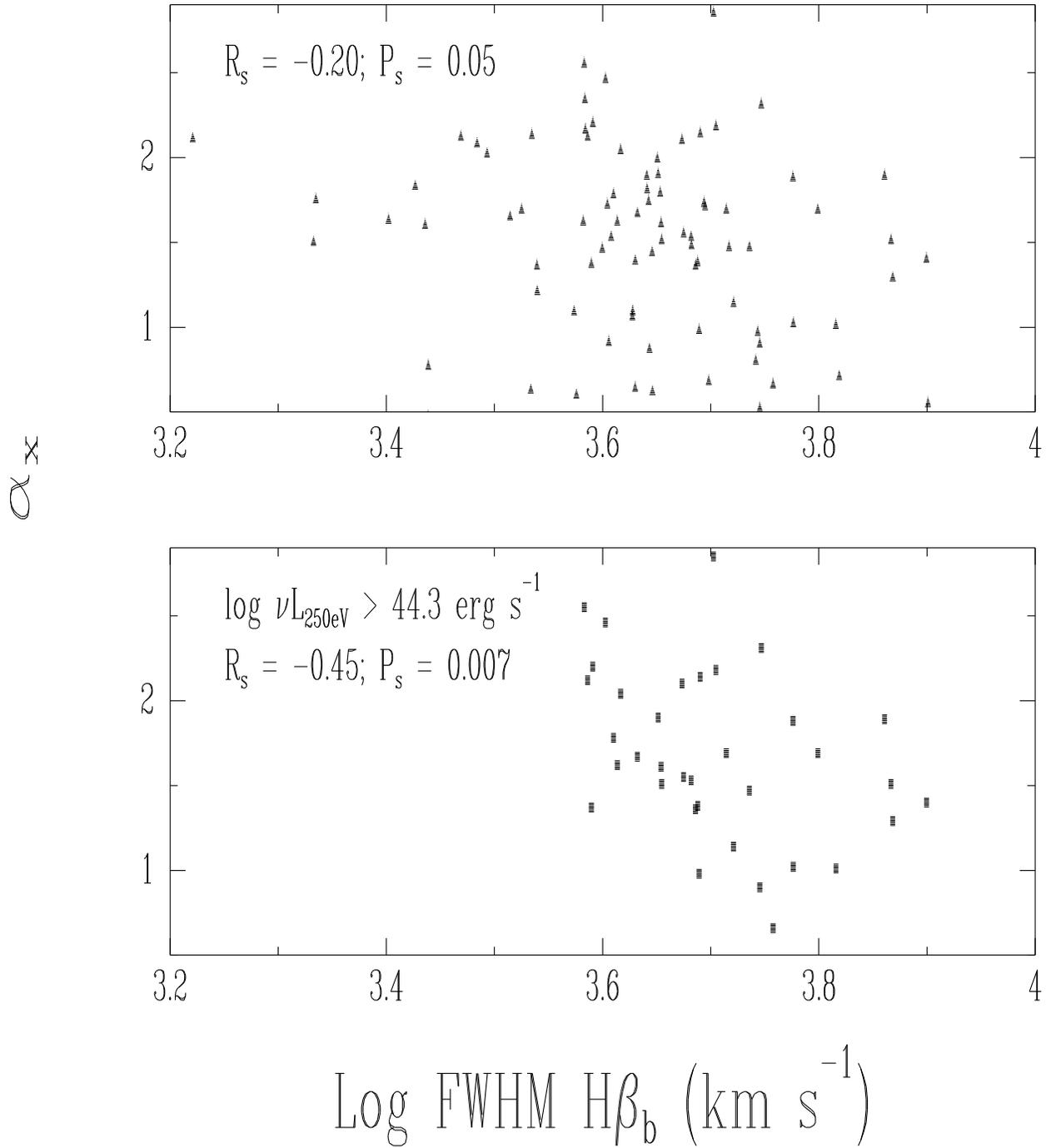}
\caption{The X-ray slope $\alpha_X$ as a function of H$\beta$ FWHM line widths.
The upper graph shows the correlation diagram for the sample, the lower graph
for the high-luminosity AGN with ${\rm Log\nu L_{250eV} > 44.3~erg~s^{-1}}$. 
\label{fig5}}
\end{figure}

\clearpage
\begin{figure}
\plotone{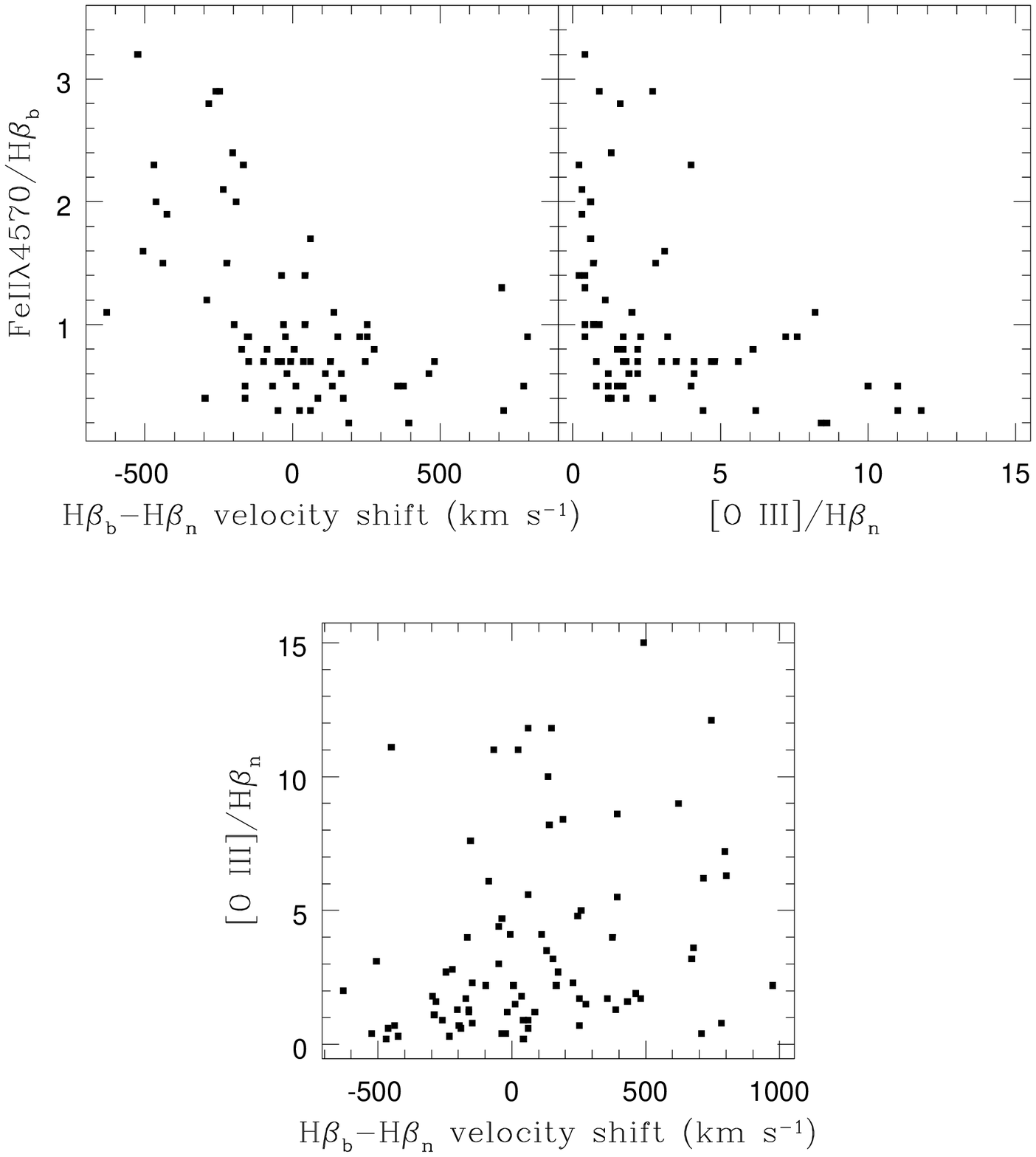}
\caption{Correlation diagrams for  Fe~II$\lambda$4570/H$\beta_b$,
H$\beta_b$-H$\beta_n$ velocity shift and [O~III]/H$\beta_n$.
\label{fig6}}
\end{figure}

\clearpage
\begin{deluxetable}{llrrrrrrrrrrrrrcc}
\rotate
\tabcolsep 1.0mm
\tablewidth{0pt}
\tablenum{1}
\tabletypesize{\scriptsize}
\tablecolumns{17}
\tablecaption{Properties of the Seyfert~1 type AGN sample}
\tablehead{
\colhead{Name} & z&\colhead{H$\beta_b$} &\colhead{H$\beta_n$} &\colhead{[O~III]} &\colhead{H$\beta_b$} &
\colhead{Fe~II} &\colhead{[O~III]} & \multicolumn{2}{c}{[O~III]/}
&\colhead{Fe~II/} &\colhead{$\Delta v$\tablenotemark{a}} &\colhead{$\Delta v$\tablenotemark{b}} &\colhead{$\Delta v$\tablenotemark{c}} &\colhead{$\alpha_{opt}$}
&\colhead{$\alpha_X$} &\colhead{$log(\nu L_{1/4}$)}
\\
\colhead{} & &\multicolumn{3}{c}{FWHM} &\colhead{EW} &\colhead{EW} &\colhead{EW} &
\colhead{H$\beta_b$} &\colhead{H$\beta_n$}
&\colhead{H$\beta_b$} &\colhead{} &\colhead{} &\colhead{} &\colhead{} &\colhead{} &\colhead{}
\\
\colhead{} & & \multicolumn{3}{c}{(km s$^{-1}$)} &\colhead{(\AA)} 
&\colhead{(\AA)} &\colhead{(\AA)} &\colhead{} &\colhead{}
&\colhead{} &\multicolumn{3}{c}{(km s$^{-1}$)} &\colhead{} 
&\colhead{} &\colhead{(erg s$^{-1}$)}
}
\startdata
1RXS J000011.9+052318&0.040 &3050  &1050 &  600  & 37.3 &26.7& 22.5 &0.6& 1.8& 0.7 &    40 & -50&  -90& 1.2&2.08$\pm$0.18 &43.6\\
1RXS J000055.5+172346&0.215 &6420  &2290 &  830  & 75.5 &64.9&  8.5 &0.1& 0.9& 0.9 &   640 & 150& -490& 0.6&1.73$\pm$0.35 &44.3\\
1RXS J000550.0+102249&0.095 &4240  & -   &  660  & 58.6 & 0.0& 45.8 &0.8&  - & 0.0 &   380 & 340&  -40& 2.0&1.06$\pm$0.47 &43.4\\
1RXS J000915.0+243758&0.118 &3880  &1540 &  640  & 34.0 &13.5& 12.9 &0.4& 2.3& 0.4 &   -70 &-150&  -80& 1.4&1.60$\pm$0.43 &44.0\\
1RXS J000953.0+371906&0.154 &4240  &   - & 1380  & 27.6 &34.6& 22.8 &0.8&  - & 1.3 &    -  &  70&    -& -  &1.84$\pm$0.46 &44.3\\
1RXS J001121.2+373919&0.104 &3770  &1010 &  860  & 52.7 &38.2& 12.4 &0.2& 1.7& 0.7 &   480 & 290& -190& -  &0.60$\pm$0.49 &43.5\\
1RXS J001452.7+270549&0.086 &3350  & 870 &  760  & 28.7 &25.5& 24.4 &0.8& 2.3& 0.9 &   230 & 200&  -30& 1.1&1.69$\pm$0.23 &43.9\\
1RXS J002007.2+324423&0.082 &3820  &1380 &  950  & 38.1 &78.7&  3.6 &0.1& 0.3& 2.1 &  -230 &-260&  -30& -  &1.62$\pm$0.45 &43.6\\
1RXS J002445.3+082047&0.067 &1660  & 690 &  690  & 31.0 &72.2&  6.2 &0.2& 4.0& 2.3 &  -170 &-220&  -50& -  &2.11$\pm$0.50 &43.3\\
1RXS J002811.6+310342\tablenotemark{d}&0.500&   - &  - &762  &  -  & -  &  8.8 & - &  - &  -  &    -  &  - &   - & 2.1&1.32$\pm$0.25 &45.5\\
1RXS J002902.7+195717&0.264 &5720  & 740 &  720  & 71.0 &24.7& 14.4 &0.2& 6.2& 0.3 &   720 & 260& -460& -  &0.66$\pm$0.63 &44.2\\
1RXS J003121.2+301558&0.200 &3940  & 715 &  630  & 23.8 &27.2& 33.1 &1.4& 9.9& 1.1 &  -370 &-310&   60& -  &1.99$\pm$0.52 &44.4\\
1RXS J003221.1+242359&0.066 &3460  &1180 &  620  & 13.2 &38.2& 14.1 &1.1& 0.9& 2.9 &  -260 &-230&   30& -  &1.21$\pm$0.33 &43.1\\
1RXS J004459.4+192143&0.179 &3900  &1380 &  750  & 20.3 &64.6&  9.5 &0.5& 0.4& 3.2 &  -520 &-470&   50& 0.8&2.20$\pm$0.24 &44.6\\
1RXS J005034.2+044152&0.097 &1780\tablenotemark{e}  &  -  &    -  &    - &   -& - & - &  - & -   &   - &   -&    -& -  &1.88$\pm$0.40 &43.5\\
1RXS J005050.6+353645&0.058 &5470  &2300 & 1230 &  48.1 &27.3& 10.8 &0.2& 0.5& 0.6 &   160 &   0& -160& 0.7&2.21$\pm$0.15 &44.0\\
1RXS J005607.6+254802&0.153 &6600  &1960 & 1230 &  73.3 &61.9& 13.8 &0.2& 0.9& 0.8 &  -760 &  90&  850& 1.1&1.71$\pm$0.39 &44.4\\
1RXS J011655.1+254937&0.099 &4000  & 910 & 1070 &  56.4 &40.1& 43.1 &0.8& 3.0& 0.7 &   -50 &-110&  -60& 0.0&- &-\\
1RXS J011745.5+363718&0.107 &3770  &1720 &  700 &  68.3 &51.8& 17.2 &0.3& 0.6& 0.8 &    40 & 230&  190& 0.7&1.03$\pm$0.28 &44.0\\
1RXS J011848.2+383626&0.216 &5040  &1590 & 1070 &  68.0 &50.0& 13.5 &0.2& 0.8& 0.7 &  -150 &  40&  190& -  &2.85$\pm$0.43 &44.8\\
1RXS J012556.6+351039&0.312 &6790  & -   &  570 &  78.3 &21.0& 12.6 &0.2&  - & 0.3 &    -  &   0&    -& -  &0.86$\pm$0.89 &44.3\\
1RXS J012923.0+223100&0.231 &4730  & 980 &  800 &  65.7 &39.4& 17.3 &0.3& 2.2& 0.6 &   170 &   0& -170& -  &1.55$\pm$0.71 &44.4\\
1RXS J013738.5+852414&0.499 &4270  &  -  &  710 &   -   & -  &  8.0 &0.2& -  & -   &    -  & 180&   - & -  &0.80$\pm$0.92 &44.8\\
1RXS J014023.0+240255&0.072 &4620  &1300 & 1260 &  48.3 &44.0& 13.7 &0.3& 0.9& 0.9 &  -200 & -70&  130& 1.6&1.89$\pm$0.37 &44.0\\
1RXS J014025.2+382400&0.067 &3110  & 800 &  620 &  72.2 &38.2& 22.6 &0.3& 4.0& 0.5 &   380 & 290&  -90& 1.7&2.02$\pm$0.24 &44.0\\
1RXS J015536.7+311525&0.135 &3830  &1750 &  910 &  24.8 &42.6&  9.3 &0.4& 0.6& 1.7 &    60 & 270&  200& 0.5&2.34$\pm$0.50 &44.1\\
1RXS J015546.4+071902&0.069 &3750\tablenotemark{e}  &1090\tablenotemark{e} &  800 &   -   & -  & 26.7 & - & -  & -   &     - &   -&    -& 0.0&1.22$\pm$0.54 &43.2\\
1RXS J015555.6+040620&0.136 &6980  &2030 &  750 &  51.9 &63.1&  9.9 &0.2& 1.1& 1.2 &     0 &   0&    0& -  &1.15$\pm$0.36 &43.8\\
1RXS J015621.6+241838&0.155 &3770  & 940 &  660 &  51.5 &70.7& 18.9 &0.4& 2.7& 1.4 &   -60 & -20&   40& -  &1.73$\pm$0.85 &44.3\\
1RXS J023601.9+082124\tablenotemark{f}&0.228 &4390  &1250 & 1230 &  82.1 &74.2& 21.2 &0.3& 1.7& 0.9 & 250 &80&-170& -  &2.86$\pm$0.62 &45.5\\
1RXS J023739.9+181945&0.168 &4560  & 950 &  850 &  87.4 &58.1& 15.8 &0.2& 5.1& 0.7 &   160 &  30& -130& 0.8&-    &-   \\
1RXS J024214.5+074442&0.451 &4950  &1410 &  700 &  59.2 & 0.0& 17.2 &0.3& 0.9& 0.0 &    60 &   0&  -60& -  &1.71$\pm$0.78 &45.5\\
1RXS J033522.6+190739&0.189 &6180  & -   &  800 &  30.8 & 5.8& 19.3 &0.6&  - & 0.2 &    -  & 200&    -& 0.9&0.20$\pm$0.45 &44.2\\
1RXS J034308.0+185836&0.110 &3420  &1300 &  830 &  75.0 &56.5& 44.4 &0.6& 1.5& 0.8 &   280 & 410&  130& 1.6&0.63$\pm$0.62 &43.8\\
1RXS J035147.6+103618&0.185 &5100  &1040 &  910 &  70.1 & 0.0& 28.9 &0.4& 7.3& 0.0 &   580 & 460& -120& 1.9&2.54$\pm$0.66 &45.2\\
1RXS J035345.9+195825&0.029 &3800  &1150 &  710 &  23.0 &14.0&  6.4 &0.3& 1.2& 0.6 &     0 &   0&    0& -  &1.73$\pm$0.72 &43.2\\
1RXS J040513.2+173647&0.098 &3850  &-    &  920 &  67.9 & -  &  9.0 &0.1& -  & -   &     - &-310&    -& -  &-    &-\\
1RXS J041147.1+132417&0.277 &2960  &1010 &  790 &  49.1 &49.5&  8.8 &0.2& 0.4& 1.0 &   -30 & 170&  200& -  &3.10$\pm$1.07 &45.8\\
1RXS J044428.6+122113&0.089 &3140  &1320 & 1280 &  33.8&106.3& 30.5 &0.9& 1.4& 3.1 &  -490 &-250&  240& 1.2&1.33$\pm$0.54 &44.0\\
1RXS J062334.6+644542&0.086 &4400  &1550 &  990 &  53.0&  0.0&117.3 &2.2& 3.6& 0.0 &   680 & 490& -190& 2.0&0.87$\pm$0.38 &43.8\\
1RXS J065136.5+563732&0.286 &4810  &1320 &  670 &  52.0& 55.5& 13.8 &0.3& 2.0& 1.1 &  -630 &-680&  -50& -  &1.53$\pm$0.59 &44.6\\
1RXS J065211.3+553436&0.064 &2750  &1010 &  620 &  25.8& 40.3& 39.4 &1.5& 3.1& 1.6 &  -510 &-400&  110& -  &0.47$\pm$0.46 &43.1\\
1RXS J071006.0+500243&0.154 &4030  &1600 &  620 &  43.7& 60.4&  7.3 &0.2& 0.2& 1.4 &    40 & 250&  210& 0.0&0.91$\pm$0.58 &44.1\\
1RXS J071339.7+382043&0.123 &4140  &1200 &  700 &  58.0&102.5& 17.0 &0.3& 0.7& 1.8 &  -490 &-340&  150& 0.5&1.33$\pm$0.46 &44.0\\
1RXS J071858.2+705920&0.066 &4810  &1310 &  850 &  20.3& 57.7& 38.2 &1.9& 1.6& 2.8 &  -280 &-250&   30& -  &1.48$\pm$0.32 &43.2\\
1RXS J071923.7+570840&0.175 &6540  &1150 &  960 & 102.7&  0.0& 81.4 &0.8& 3.2& 0.0 &   670 & 560& -110& 0.2&1.01$\pm$0.33 &44.3\\
1RXS J072220.6+732041&0.350 &3570  &1320 &  820 &  59.7& 76.4& 10.6 &0.2& 0.7& 1.3 &  -150 & 300&  450& -  &2.71$\pm$0.35 &45.1\\
1RXS J072408.4+355309&0.138 &7260  & 820 &  680 &  94.4&  0.0& 74.0 &0.8&11.1& 0.0 &  -450 &-510&  -60& 0.6&1.89$\pm$0.18 &44.7\\
1RXS J072826.6+303407&0.100 &3980  &1010 &  600 &  59.7& 48.1& 15.3 &0.3& 2.2& 0.8 &     0 &   0&    0& 1.4&1.46$\pm$0.53 &43.7\\
1RXS J072912.9+410552&0.103 &4460  &1580 & 1040 &  47.6& 41.6& 29.4 &0.6& 3.7& 0.9 &  -230 & 190&  420& -  &1.28$\pm$0.39 &43.8\\
1RXS J073308.7+455511&0.142 &8070  &3240 &  930 &  58.1& 48.9& 16.7 &0.3& 0.3& 0.8 &  -440 &-190&  250& 0.1&1.28$\pm$0.29 &44.5\\
1RXS J073735.8+430856&0.119 &4500  & 670 &  660 &  33.8& -   & 51.7 &1.5&14.9& -   &     0 &   0&    0& -  &-    &-\\
1RXS J074228.1+465648&0.170 &4050  &  -  &  620 &  99.3& 69.3& 25.4 &0.3& -  & 0.7 &   -   & 330&    -& 1.4&-    &-\\
1RXS J074311.3+742926&0.332 &7380  &1060 &  840 &  94.0& 32.1& 31.4 &0.3& 4.4& 0.3 &   -50 &-140&  -90& -  &1.29$\pm$0.26 &44.9\\
1RXS J074759.2+205229&0.156 &5540  & 820 &  690 &  71.7& 48.7& 12.4 &0.2& 3.5& 0.7 &   130 &  70&  -60& 1.2&0.97$\pm$0.45 &44.0\\
1RXS J074906.2+451039&0.192 &4500  & 800 &  800 & 100.4& 55.1& 80.4 &0.8& 8.4& 0.6 &     0 &   0&    0& 0.4&0.04$\pm$0.84 &43.6\\
1RXS J075026.6+270726&0.154 & -    & -   &  -   & -    & -   & -    &-  & -  & -   &     - &   -&    -& -   &1.01$\pm$0.38& 43.7\\
1RXS J075039.5+624205&0.194 &6300  &1330 & 1500 &  58.5& 18.0& 22.3 &0.4&11.0& 0.3 &    20 & -90& -110& 0.9 &1.69$\pm$0.36& 44.4\\
1RXS J075408.7+431611&0.350 &-     & -   &  570 &  -   &  -  & 31.5 &-  & -  & -   &     - &   -&    -& -   &1.38$\pm$0.30& 45.0\\
1RXS J075409.5+512436&0.331 &3680  & -   &  730 &  99.0&108.2& 11.0 &0.1& -  & 1.1 &     - &   0&    -& -   &-&- \\
1RXS J075634.7+362232\tablenotemark{g}&0.884&-&-  &    - &  -   &  -  &   -  &-  & -  & -   &     - &   -& -& - &1.60$\pm$0.42& 45.8\\
1RXS J075910.0+115152&0.050 &4470 & 1180 &  700 &  25.9& 52.8& 12.7 &0.5& 0.6& 2.0 &  -190 &-190&    0& 1.4 &1.99$\pm$0.32& 43.3\\
1RXS J080132.3+473618&0.157 &7840 &  -   &  900 & 119.0& 69.7& 13.1 &0.1& -  & 0.6 &  -530 &-190&  340& 0.4 &1.29$\pm$0.17& 44.6\\
1RXS J080358.9+433248&0.451 &5580 & 1340 &  670 &  75.6&  0.0& 38.4 &0.5& 1.3& 0.0 &   390 & 570&  180& -   &2.31$\pm$0.32& 45.3\\
1RXS J080534.6+543132&0.406 &5780 & 1530 &  730 &  51.5&  -  & 30.1 &0.6& 1.5& -   &  -370 &-250&  120& -   &2.01$\pm$0.22& 45.5\\
1RXS J080548.5+112617&0.144 &4480 &  880 &  770 &  89.6&  0.0& 50.0 &0.6& 5.0& 0.0 &   260 &   0& -260& 0.7 &1.90$\pm$0.22& 44.2\\
1RXS J080639.9+724831&0.099 &1760 &  740 &  820 &  13.7& 34.4& 61.9 &4.5& 6.2& 2.5 &  -190 &-220&  -30& 0.7 &1.25$\pm$0.34& 43.7\\
1RXS J080727.2+140534&0.074 &2450 &  -   &  -   &  51.7& 70.2&$<$20&$<$0.4& -&1.4 &    -  &   0&    -& -   &0.98$\pm$0.45& 43.3\\
1RXS J081237.9+435636&0.183 &4110 &  -   &  700 &  51.5& 21.6& 11.8 &0.2& -   &0.4 &    -  & -90&    -& 0.4 &1.14$\pm$0.44& 44.2\\
1RXS J081502.4+525304&0.126 &4110 & 1080 &  770 &  75.8& 41.2& 14.7 &0.2& 1.7 &0.5 &   360 & 230& -130& 0.2 &1.62$\pm$0.24& 44.2\\
1RXS J081651.4+180252&0.158 &3890 &  810 &  700 &  68.4&  0.0& 84.6 &1.2& 9.0 &0.0 &   620 & 680&   60& 0.9 &1.37$\pm$0.20& 44.5\\
1RXS J081718.6+520200&0.040&2170\tablenotemark{e} &  470 &  -   &  -   & 25.2&  -   & - &  -  & - & - & -& -& - &1.09$\pm$0.42& 42.6\\
1RXS J081738.0+242333&0.283 &5560 &  960 &  740 &  73.0& 14.9&125.4 &1.7& 8.6 &0.2 &   390 & 350&  -40& -   &0.90$\pm$0.74& 44.3\\
1RXS J082633.1+074259&0.312 &3460 & 1010 &  880 &  50.0& 65.6&  6.0 &0.1& 1.3 &1.3 &  -130 & 310&  440& -   &-   & -    \\
1RXS J082942.2+415442&0.126 &7960 &  800 &  910 &  70.1&  0.0& 98.6 &1.4&12.1 &0.0 &   750 & 700&  -50& 1.8 &0.55$\pm$0.39& 43.5\\
1RXS J083737.3+254746&0.080 &-    &  -   &  960 &   0.0&   - & 13.3 &-  & -   &-   &     - &   -&    -& -   &-   & -    \\
1RXS J083820.6+241025&0.043 &4980 &  780 &  750 &  26.0& 18.0& 11.0 &0.0& 3.0 &0.0 &  -680 &-770&  -90& -   &1.70$\pm$0.29& 43.2\\
1RXS J084026.2+033316&0.060 &4940 & 1510 &  990 &  55.6&  0.0& 28.5 &0.5& 1.6 &0.0 &   430 & 390&  -40& 0.3 &1.73$\pm$0.21& 43.6\\
1RXS J084302.0+030226&0.512 &1920 &  -   &  -   &  53.1& 43.3&  -   &-  & -   &0.8 &     - &   -&    -& -   &2.25$\pm$0.38& 45.4\\
1RXS J084622.3+031318&0.107 &4380 & 1000 &  960 &  71.6& 49.7& 24.2 &0.3& 4.1 &0.7 &     0 &-150& -150& 0.9 &1.81$\pm$0.26& 44.1\\
1RXS J084856.8+584150&0.134 &-    & -    &  890 &  -   &  0.0& 23.3 &-  & -   &-   &     - &   -&    -& -   &1.29$\pm$0.47& 43.7\\
1RXS J084918.7+531806&0.112 &2750 &  810 &  990 &  23.3& 20.6& 29.7 &1.3& 7.6 &0.9 &  -150 & -20&  130& 1.5 &0.77$\pm$0.48& 43.2\\
1RXS J085233.3+422539&0.137 &2670 &  750 &  700 &  14.8& 43.3& 22.8 &1.5& 2.7 &2.9 &  -250 &-220&   30& 1.1 &1.83$\pm$0.38& 43.9\\
1RXS J085358.8+770054&0.106 &5480 & 2300 &  620 &  23.1& 40.8& 11.7 &0.5& 0.8 &1.8 &  -640 &-540&  100& 1.6 &1.34$\pm$0.16& 43.8\\
1RXS J085443.8+401912&0.152 &5520 & 1100 &  810 &  60.6& 37.1& 19.7 &0.3& 1.9 &0.6 &   460 & 440&  -20& 1.8 &0.80$\pm$0.30& 43.7\\
1RXS J085706.2+621050&0.160 &   - &    - &    - &     -&    -&    - &  -&   - &  - &     - &   -&    -& -   &0.08$\pm$0.97& 43.4\\
1RXS J085820.6+642100&0.115&3060\tablenotemark{e} & 1260\tablenotemark{e} &970 &-&-&63.7 &-&- &- &- &-&-& - &1.18$\pm$0.40& 43.6\\
1RXS J085902.0+484611&0.083 &4420 & 1300 &  910 &  80.0& 72.2& 36.1 &0.5& 2.3 &0.9 &  -150 &-130&   20& 0.6 &1.44$\pm$0.11& 44.1\\
1RXS J190910.3+665222&0.191 &3830 &  600 &  670 &  97.0&111.7&  7.0 &0.1& 1.1 &1.2 &  -290 & -40&  250& 0.9 &2.55$\pm$0.12& 44.8\\
1RXS J193929.8+700752&0.116 &4850 & 1690 & 1160 &  60.9& 54.1& 14.1 &0.2& 0.4 &0.9 &   -20 &  80&  100& 0.2 &1.36$\pm$0.15& 44.4\\
1RXS J210550.5+041435&0.139 &   -  &   - &    - &     -&    -&    - &  -&   - &  - &     - &   -&    -& -   &1.54$\pm$0.54& 43.9\\
1RXS J212654.3+112403&0.096 &2730  &1600 & 1120 &  66.7& 33.3& 14.1 &0.2& 0.8 &0.5 &   780 & 350& -430& 0.7 &1.60$\pm$0.28& 44.1\\
1RXS J214309.6+153658&0.385 &7610    & - & 1230 &  72.7&    0& 10.9 &0.1& -   &  0 &     - & 120&    -& -   &1.00$\pm$0.94& 44.9\\
1RXS J214655.2+092102\tablenotemark{g}&0.984&-& - &  -   &  -   &    -&    - &  -& -   &-   &     - &  - &   - &  -  &0.83$\pm$0.61& 45.6\\
1RXS J215912.7+095247&0.102 &5380 &    - & 1110 &  71.4& 26.0& 50.0 &0.7& -   &0.4 &     - &  0 &   - & 1.3 &1.51$\pm$0.25& 44.2\\
1RXS J221537.8+290239&0.229 &5980 &  600 &  640 &  70.0&  0.0& 54.5 &0.8&11.8 &0.0 &   150 & 220&   70& -   &1.02$\pm$0.36& 44.4\\
1RXS J221918.8+120757&0.082&2080\tablenotemark{e} &    - &    - &     -&    -&    - &-  & -   &-   &- &-&-& - &2.58$\pm$0.22& 44.4\\
1RXS J222537.5+225917&0.173 &2150 & 1100 & 1100 &  52.5& 39.4& 17.8 &0.3& 4.7 &0.7 &   -40 &  0 &   40& 0.6 &1.50$\pm$0.47& 44.0\\
1RXS J222846.3+333500&0.090 &3270 &  900 & 1260 &  44.4& 24.1& 46.1 &1.0&11.0 &0.5 &   -70 &   0&   70& -   &1.65$\pm$0.73& 44.0\\
1RXS J222934.3+305720&0.319 &7930 & 1390 & 1110 &  88.0&  0.0& 30.0 &0.3& 6.3 &0.0 &   800 & 900&  100& -   &1.40$\pm$0.43& 44.8\\
1RXS J223019.5+163114&0.084 &4270 & 1740 & 1310 &  55.8& 34.9& 17.2 &0.3& 1.2 &0.6 &   -20 &  80&  100& 0.8 &1.39$\pm$0.31& 43.8\\
1RXS J224621.7+314206&0.145 &   - &    - &  660 &     0&   - & 25.3 &-  & -   &-   &   -   &  - &   - & -   &1.48$\pm$0.30& 44.4\\
1RXS J224939.6+110016&0.084 &4390 & 1440 & 1220 &  24.7& 47.2&  3.5 &0.1& 0.3 &1.9 &  -430 &-180&  250& 0.9 &1.74$\pm$0.25& 44.0\\
1RXS J225148.5+341937&0.132 &3050 & 1120 &  590 &  51.4& 55.5& 10.5 &0.2& 0.6 &1.1 &    40 & 220&  180& 0.6 &1.43$\pm$0.22& 44.5\\
1RXS J225209.6+264236&0.069 &2160 &  700 &  700 &  16.7& 39.4& 15.5 &0.9& 1.3 &2.4 &  -200 &-200&    0& 1.3 &1.75$\pm$0.26& 43.7\\
1RXS J225639.4+261846&0.176 &   - &    - &    - &     -&    -&    - &  -&   - &  - &     - &   -&    -& -   &0.87$\pm$0.79& 43.8\\
1RXS J230328.0+144341&0.080 &   - &    - &    - &     -&    -&    - &  -&   - &  - &     - &   -&    -& -   &0.20$\pm$0.46& 42.9\\
1RXS J231456.0+224325&0.171 &3750 & 1230 & 1170 &  62.8&127.8& 22.4 &0.4& 0.7 &2.0 &  -360 &-360&    0& 0.9 &0.98$\pm$0.35& 44.1\\
1RXS J231517.5+182825&0.104 &3310 & 1190 & 1200 &  25.0& 18.3& 10.7 &0.4& 6.4 &0.7 &     0 &   0&    0& -   &-&- \\
1RXS J232554.6+215310&0.120 &5260 &  950 &  770 & 125.2&  0.0& 62.5 &0.5& 5.5 &0.0 &   390 & 330&  -60& 1.1 &1.14$\pm$0.13& 44.5\\
1RXS J232841.4+224853&0.129 &5400 &  -   &  810 & 121.7&  0.0& 34.6 &0.3& -   &0.0 &     - &   0&    -& 1.5 &1.10$\pm$0.54& 43.7\\
1RXS J234728.8+242743&0.159 &4270 & 2550 & 1350 & 105.1&  0.0& 75.2 &0.7& 2.2 &0.0 &   980 & 770& -210& 1.7 &0.64$\pm$0.17& 44.2\\
1RXS J234734.6+271910\tablenotemark{g}&0.646&- &  - &    - &     -&   - &    - &  -&   - &  - & - & -& -& -   &-&- \\
1RXS J235152.7+261938&0.040 &3750 & 1850 & 1460 &  47.7& 19.0& 36.7 &0.8& 1.8 &0.4 &  -300 &-250&   50& 0.9 &1.09$\pm$0.21& 43.3\\
1RXS J235754.3+132418&0.115 &4290 & 2000 & 1480 & 102.3&101.9& 35.5 &0.3& 0.9 &1.0 &    40 & 380&  340& 1.5 &1.67$\pm$0.18& 44.2\\
III ZW 2             &0.090 &4990 &  700 &  630 &  72.1& 33.3& 47.1 &0.7&10.0 &0.5 &   140 &   0& -140& 0.5 &0.68$\pm$0.24& 44.0\\
4C 25.01             &0.284 &4310 & 1050 &  840 &  74.6& 79.1& 38.8 &0.5& 8.2 &1.1 &   140 & 140&    0&   - &1.51$\pm$0.43& 44.9\\
PG 0026+12           &0.142 &4890 & 1120 &  690 &  36.1& 19.2& 14.8 &0.4& 1.5 &0.5 &     0 & 360&  360& 0.9 &0.98$\pm$0.13& 44.5\\
S 10785              &0.134 &5970 & 1290 &  850 &  89.1& 65.4& 26.2 &0.3& 5.6 &0.7 &    60 & 260&  200& 1.5 &1.88$\pm$0.21& 44.6\\
MARK 1148            &0.064 &5560 &  840 &  800 & 135.3& 22.8& 55.0 &0.4& 8.4 &0.2 &   190 & 290&  100& 1.2 &0.52$\pm$0.16& 43.5\\
PG 0052+251          &0.155 &5110 &  780 &  780 &  68.8& 19.2& 30.0 &0.4& 8.6 &0.3 &    60 &  20&  -40& 0.6 &1.38$\pm$0.18& 44.9\\
B2 0110+29           &0.363 &7360 &  680 &  750 &  77.6&  0.0& 63.1 &0.8&15.0 &0.0 &   490 & 520&   30& -   &1.51$\pm$0.63& 44.8\\
3C 48.0              &0.367 &4080 & 1840 & 1200 &  31.0& 46.2& 37.2 &1.2& 2.8 &1.5 &  -220 & 270&  490& 0.3 &1.78$\pm$0.11& 45.9\\
MS 02448+1928        &0.176 &4510 & 2000 &  780 &  70.0& 59.5& 23.9 &0.3& 1.1 &0.9 &   780 & 720&  -60& -   &1.93$\pm$0.77& 44.7\\
MS 03574+1046        &0.182 &4160 & 1660 &  890 &  42.2& 72.2& 13.0 &0.3& 0.5 &1.7 &  -380 & -90&  290& 1.1 &2.73$\pm$1.10& 45.3\\
RX J04506+0642       &0.118 &3460 & 1150 &  940 &  22.5& 45.2& 11.0 &0.5& 0.6 &2.0 &  -460 &-360&  100& 0.7 &1.36$\pm$1.11& 44.0\\
MS 07007+6338        &0.152 &5070 & 2070 &  840 &  49.8& 63.6&  8.4 &0.2& 0.4 &1.3 &   710 & 450& -260& 0.2 &2.18$\pm$0.33& 44.4\\
VII ZW 118           &0.079 &4910 & 2300 &  860 &  58.1& 71.4& 14.3 &0.2& 0.5 &1.2 &   490 & 680&  190& 0.5 &1.95$\pm$0.09& 44.7\\
F07144+4410          &0.061 &3820 & 1380 &  830 &  95.8& 65.3&105.1 &0.8& 2.4 &0.7 &  -120 & -70&   50& 0.7 &1.16$\pm$0.21& 44.1\\
MS 07199+7100        &0.125 &4020 & 1330 &  680 &  54.8& 50.9& 26.3 &0.5& 7.2 &0.9 &   800 & 230& -570& -   &1.72$\pm$0.23& 44.1\\
RX J07491+2842       &0.345 &4710 &  820 &  620 &  84.3& 57.7& 12.3 &0.1& 4.8 &0.7 &   250 & 140& -100& -   &2.10$\pm$0.39& 44.7\\
RX J07527+2617       &0.082 &2940 & 1060 &  780 &  38.3& 55.0&  7.9 &0.2& 0.4 &1.4 &   -40 & -40&    0& -   &2.12$\pm$0.23& 44.0\\
B3 0754+394          &0.096 &3840 & 1430 &  870 &  57.3& 46.8& 38.1 &0.7& 1.7 &0.8 &  -170 &-130&   40& 0.6 &2.16$\pm$0.33& 43.9\\
KUV 07549+4228\tablenotemark{f}&0.210 &4620 & 1520 &  780 &  57.5& 74.2& 48.5 &0.8& 2.4 &1.3 &   170 &  70& -100& 0.3 &2.14$\pm$0.22& 44.9\\
PG 0804+761          &0.100 &5490 & 2380 &  950 &  65.4& 85.1& 11.3 &0.2& 0.3 &1.3 &  -300 & -60&  240&-0.2 &1.30$\pm$0.07& 44.7\\
RX J08297+3252       &0.127 &3420 &  820 &  710 &  33.7& 30.9& 44.2 &1.3& 3.2 &0.9 &   150 & 150&  0  & 1.3 &2.13$\pm$0.42& 43.9\\
RX J08307+3405       &0.063 &4500 & 1090 &  680 &  37.7& 21.6& 36.9 &1.0& 4.1 &0.6 &   110 & 140&   30& 1.6 &1.79$\pm$0.26& 43.6\\
US 1329              &0.249 &5180 & 1280 &  760 & 112.4& 48.2& 55.8 &0.5& 2.7 &0.4 &   170 & 120&  -50& -   &1.69$\pm$0.15& 45.2\\
VII ZW 244           &0.131 &4510 & 2000 &  770 &  60.1& 88.0& 15.2 &0.3& 0.7 &1.5 &  -440 &-190&  250& 0.8 &1.61$\pm$0.17& 44.2\\
KUV 08482+4146       &0.134 &4130 &  -   &  740 &  62.2& 39.4& 11.4 &0.2&  -  &0.7 &    -  &-520&  -  & 1.2 &1.47$\pm$0.26& 44.0\\
MARK 1220            &0.064 &4050 &  770 &  930 &  30.7& 23.9& 25.8 &0.8& 6.1 &0.8 &   -90 & 140&  220& -   &1.53$\pm$0.31& 43.5\\
4C 73.18             &0.303 &5670 & 1770 &  870 &  53.6&  0.0& 10.3 &0.2& 0.6 &0.0 &   -60 &  50&  110& -   &1.02$\pm$0.16& 45.1\\
RX J20389+0303\tablenotemark{f}&0.137 &4410 & 1410 &  750 &  65.5& 67.7& 31.3 &0.3& 0.7 &1.0 &    40 &-100& -140& 1.3 &1.81$\pm$0.74& 44.1\\
S5 2116+81           &0.086 &4130 &  -   &  740 & 122.7& 22.5& 92.6 &0.8& -   &0.2 &   -   &  0 &   - & 1.8 &0.71$\pm$0.14& 44.0\\
MS 21283+0349        &0.094 &4900 & 1330 &  660 & 105.4& 50.9& 54.8 &0.5& 2.5 &0.5 &  -370 &-310&   60& 0.4 &2.14$\pm$0.22& 44.6\\
II ZW 136            &0.063 &3770 &  980 &  760 &  81.4& 72.5& 17.4 &0.2& 0.9 &0.9 &   250 & 120& -130& 0.9 &2.46$\pm$0.13& 44.7\\
MS 22102+1827        &0.079 &2950 &  870 &  900 &  27.3& 68.9&  6.2 &0.1& 0.3 &2.5 &  -120 &-120&    0& 1.3 &2.49$\pm$0.27& 44.0\\
RX J22186+0802       &0.120 &2530 &  780 &  810 &  22.5& 36.1& 23.7 &1.1& 3.5 &1.6 &  -140 &  90&  220& 1.6 &-&- \\ 
RX J22198+0935       &0.060 &2520 &  910 &  740 &  15.1&  7.9& 11.4 &0.8& 1.2 &0.5 &  -160 &-100&   60& -   &1.63$\pm$0.45& 43.1\\
PG 2233+134          &0.325 &4750 & 1340 &  740 &  41.1& 67.8& 10.4 &0.3& 0.4 &1.7 &  -570 &-410&  160& -   &0.90$\pm$0.62& 44.3\\
AKN 564              &0.025 &3860 &  970 &  720 &  31.6& 23.1& 50.6 &1.6& 2.2 &0.7 &  -100 &-150&  -50& 1.4 &2.12$\pm$0.04& 44.5\\
KUV 22497+1439       &0.135 &4140 & 1560 & 1040 &  47.5&110.1&  4.2 &0.1& 0.2 &2.3 &  -470 &   0&  470& 0.6 &2.04$\pm$0.20& 44.5\\
RX J2256.6+0525      &0.065 &4240 & 1230 &  730 &  82.9& 37.0& 53.5 &0.6& 1.3 &0.4 &  -160 & -50&  110& 1.1 &1.09$\pm$0.20& 43.8\\
KAZ 320              &0.034 &4430 & 1230 &  830 &  48.3& 17.8& 53.8 &1.1& 1.2 &0.4 &    90 & 220&  130& 1.3 &0.62$\pm$0.12& 43.1\\
RX J23171+0838       &0.165 &5440 &  910 &  690 &  85.7& 21.1& 31.1 &0.4&10.1 &0.2 &  250  & 60 &-190 & 0.7 &1.47$\pm$0.37& 44.2\\
\enddata
\tablenotetext{a}{H$\beta$ broad component velocity shift with respect to 
narrow component. A positive velocity indicates a redshift of H$\beta$ 
broad component with relative to narrow component, whereas
a negative velocity refers to a blueshift.}
\tablenotetext{b}{H$\beta$ broad component velocity shift relative to the 
systemic frame. A positive velocity corresponds to a redward shift of 
H$\beta$ broad component with respect
to systemic, while a negative velocity represents a blueshift.}
\tablenotetext{c}{H$\beta$ narrow component velocity shift relative to the 
systemic frame. A positive velocity corresponds to a redward shift of 
H$\beta$ narrow component with respect
to systemic, while a negative velocity represents a blueshift.}
\tablenotetext{d}{H$\beta$ is weak.}
\tablenotetext{e}{H$\beta$ FWHM measurements are based on the H$\alpha$.}
\tablenotetext{f}{Sources show obvious asymmetric 
[O~III]$\lambda\lambda4959,5007$ profiles}
\tablenotetext{g}{H$\beta$ and [O~III] are not in the observed 
wavelength range.}
\tablecomments{
The Prefix of the object name indicates the origin of sources. 
`1RXS~J' represents the
source that was identified as emission line AGN in our identification 
program, while other prefix refer to the known AGN that meet our 
selection criteria. The typical
uncertainty of the the emission line fluxes is about 10-20\%. 
Errors for FWHM and velocity
shift measurements are $\la$150~km~s$^{-1}$.}
\end{deluxetable}

\clearpage
\begin{deluxetable}{lllllllllllll}
\rotate
\tabletypesize{\scriptsize}
\tablecolumns{13}
\tablewidth{0pt}
\tablenum{2}
\tablecaption{Spearman Rank-Order Correlation Coefficient Matrix}
\tablehead{
\colhead{Property} &\colhead{(1)} &\colhead{(2)} &\colhead{(3)} &\colhead{(4)} &\colhead{(5)} &\colhead{(6)} &\colhead{(7)} &\colhead{(8)}
&\colhead{(9)} &\colhead{(10)} &\colhead{(11)} &\colhead{(12)}}
\startdata
(1)FWHM(H$\beta_b$)&--  &  0.002   &  0.032    &    -0.32  & 0.35   &  -0.59  &   -0.05  &   0.34 & -0.20  &  -0.04  & 0.38   &   0.48\\
              &         &          &           &   (0.0020)&(0.0009)& (0.0000)&          &  (0.001)&        &        & (0.0003)&(0.0000)\\
(2)FWHM(H$\beta_n$)&    &   --     &  0.46     &    0.30   &-0.263  &  0.23   &   -0.294 &   -0.15 & 0.23   & -0.104 & -0.66   &  0.11\\
              &         &          &  (0.0000) &   (0.004) &        &         &   (0.005)&          &        &        &(0.0000)&        \\
(3)FWHM([OIII])&       &          &   --      &    -0.03  & 0.004  & -0.02   &   -0.08  &   0.02  &-0.06    &-0.166 & -0.15   & -0.01\\
              &         &          &           &           &        &         &          &          &        &        &        &        \\
(4)EW FeII    &         &          &           &      --   & -0.59  & 0.81    &   -0.546 &  -0.471 & 0.41    &-0.129  &-0.544  & 0.06\\
              &         &          &           &           &(0.0000)& (0.0000)&  (0.0000)&  (0.0000)&(0.0001)&        &(0.0000)&        \\
(5)EW [OIII]  &         &          &           &           &  --    & -0.64   &   0.73   &   0.33   &-0.39   &0.195   &0.67    & 0.05\\
              &         &          &           &           &        & (0.0000)&  (0.0000)&  (0.0019)&(0.0002)&        &(0.0000)&        \\
(6)FeII/H$\beta_b$&     &          &           &           &        &   --    &   -0.23  &  -0.64   &0.34    &-0.055  &-0.60   & 0.24\\
              &         &          &           &           &        &         &          &  (0.0000)&(0.0011)&        &(0.0000)&        \\
(7)[OIII]/H$\beta_b$&   &          &           &           &        &         &    --    &  0.009   &-0.33   &0.27    &0.45    &-0.314\\
              &         &          &           &           &        &         &          &          &(0.0017)&        &(0.0000)&(0.0038)\\
(8)$\Delta v$\tablenotemark{a} & &        &           &           &        &         &          &    --    &-0.24   &0.147   &0.375   & 0.14\\
              &         &          &           &           &        &         &          &          &        &        &(0.0003)&        \\
(9)$\alpha_X$&          &          &           &           &        &         &          &          &--      &-0.236  &-0.37   & 0.36\\
              &         &          &           &           &        &         &          &          &        &        &(0.0004)&(0.001)\\
(10)$\alpha_{opt}$ &    &          &           &           &        &         &          &          &       &--       & 0.109  & -0.29\\
              &         &          &           &           &        &         &          &          &       &         &        &(0.0069)\\
(11)[OIII]/H$\beta_n$&  &          &           &           &        &         &          &          &       &         &    --  & 0.1\\
              &         &          &           &           &        &         &          &          &       &          &       &        \\
(12)Log~$\nu$L$_{250eV}$ &    &          &           &           &        &         &          &          &       &          &       & --\\
\enddata
\tablenotetext{a}{velocity shift between broad and narrow Gaussian components of H$\beta$ in units of km~s$^{-1}$.}
\end{deluxetable}

\end{document}